\documentclass[12pt]{article}
\usepackage{multirow}
\usepackage{newtxtext,newtxmath}

\usepackage{graphicx}

\usepackage[letterpaper,margin=1in]{geometry}

\renewenvironment{abstract}
	{\quotation}
	{\endquotation}

\date{}

\makeatletter
\renewcommand{\fnum@figure}{\textbf{Figure \thefigure}}
\renewcommand{\fnum@table}{\textbf{Table \thetable}}
\makeatother

\usepackage{scicite}

\usepackage{url}

\def\scititle{
	Restoring Network Evolution from Static Structure
}
\title{\bfseries \boldmath \scititle}

\author{
	Jiu Zhang$^{1}$,
	Zhanwei Du$^{2}$,
	Hongwei Hu$^{1}$,
    Ke Wu$^{3}$,
    Tongchao Li$^{4}$,
    \and
    Chuan Shi$^{5}$,
    Xiaohui Huang$^{1}$,
    Yamir Moreno$^{6,7}$,
    Yanqing Hu$^{1,8\ast}$
    \and
	\small$^{1}$Department of Statistics and Data Science, College of Science, \and 
    \small Southern University of Science and Technology, Shenzhen 518055, China.\and
	\small$^{2}$School of Public Health, Li Ka Shing Faculty of Medicine,  University of Hong Kong,\and
    \small Hong Kong Special Administrative Region of China 999077, China.\and
    \small$^{3}$Institute of Risk Analysis, Prediction and Management, Academy for Advanced Interdisciplinary Sciences, \and
    \small Southern University of Science and Technology, Shenzhen 518055, China.\and
    \small$^{4}$Liangzhu Laboratory, MOE Frontier Science Center for Brain Science and Brain-machine Integration, \and
    \small State Key Laboratory of Brain-machine Intelligence, Zhejiang University, Hangzhou 311121, China\and
    \small$^{5}$School of Computer Science, Beijing University of Posts and Telecommunications, Beijing 100876, China\and
    \small$^{6}$Institute for Biocomputation and Physics of Complex Systems (BIFI), \and
    \small Universidad de Zaragoza, 50018 Zaragoza, Spain\and
    \small$^{7}$Department of Theoretical Physics, University of Zaragoza, 50009 Zaragoza, Spain \and
    \small$^{8}$Center for Complex Flows and Soft Matter Research, \and
    \small Southern University of Science and Technology, Shenzhen 518055, China\and
	\small$^\ast$Corresponding author. Email: yanqing.hu.sc@qq.com\and
}

\begin{document} 

\maketitle


\newpage
\begin{abstract} \bfseries \boldmath

The dynamical evolution of complex networks underpins the structure–function relationships in natural and artificial systems. Yet, restoring a network’s formation from a single static snapshot remains challenging. Here, we present a transferable machine-learning framework that infers network evolutionary trajectories solely from present topology. By integrating graph neural networks with transformers, our approach unlocks a latent temporal dimension directly from the static topology. Evaluated across diverse domains, the framework achieves high transfer accuracy of up to 95.3\%, demonstrating its robustness and transferability. Applied to the \emph{Drosophila} brain connectome, it restores the formation times of over 2.6 million neural connections, revealing that early-forming links support essential behaviors such as mating and foraging, whereas later-forming connections underpin complex sensory and social functions. These results demonstrate that a substantial fraction of evolutionary information is encoded within static network architecture, offering a powerful, general tool for elucidating the hidden temporal dynamics of complex systems.
\end{abstract}

\newpage
\noindent
Complex networks provide a universal framework for modeling interactions within complex systems \cite{str01,alb02,boc06}, with broad applications in biology \cite{scm05,huj16,lin24}, ecology \cite{mon06,gjx16,bod19}, and social sciences \cite{pal05,wds13,fos18}. These networks inherently evolve, their topologies transitioning from simple to intricate structures under sustained natural or social pressures, continually adapting to dynamic environments. 
Restoring a network’s entire evolutionary trajectory from its current topology is thus of profound theoretical and practical significance. Such inference can uncover the mechanisms governing network evolution, illuminate the intrinsic connections between structure and function, and facilitate predictions of future dynamics.

The complexity inherent in the evolution and structure of networked systems restricts the ability of existing models to fully explain their dynamics. For example, the well-known preferential attachment (PA) mechanism \cite{bar99,eie03} successfully accounts for scale-free degree distributions in real-world networks, yet it falls short in addressing community structures \cite{wat98,for10}. More recent efforts have employed supervised machine-learning methods trained on partially time-resolved data to restore growth histories with encouraging accuracy \cite{wan24}. However, such approaches rely on temporal information that is often unavailable in practice, leaving most empirical networks accessible only through static snapshots. The absence of general, data-efficient methods to infer temporal evolution from static structure has thus remained a central limitation in network science.

An alternative avenue lies in exploiting cross-network regularities. If the evolutionary principles captured in one system are transferable to another, models could infer growth histories even in the absence of explicit time-stamped data. This capacity would unlock broad applications: tracing the developmental trajectories of neural connectomes to probe the origins of intelligence \cite{rot05}, restoring the evolution of protein--protein interaction networks to elucidate cellular organization and mechanisms \cite{bar04}, and analyzing the evolution of collaboration networks to aid law enforcement efforts \cite{lop22}.

Here, we introduce a transferable machine-learning framework that restores the evolutionary trajectories of complex networks using only their final structures. The framework integrates the local message-passing capabilities of graph neural networks (GNNs) \cite{spa97,ktn17,ham17} with the global contextual modeling power of transformers \cite{vas17,vpd21,ycx21}. By combining local and global structural information, the model leverages the principle that a substantial fraction of a network’s dynamic history is implicitly encoded in its current static topology. 
Validated across biological, ecological, technological, and social systems, the framework demonstrates robust cross-network transferability, achieving transfer accuracy up to 95.3\% and generalizing to unseen targets without requiring temporal data.
As a proof of concept, we restore the formation timeline of approximately 2.6 million neural connections across 140,000 neurons in the \emph{Drosophila} brain, revealing a clear correspondence between predicted formation time and functional essentiality. Detailed examination of the olfactory circuit’s evolution indicates that connections associated with essential functions, such as mating and foraging, emerge earlier, while those linked to complex social behaviors and specific odor perception develop later. Together, these findings show that much of a network’s evolutionary information is preserved in its static architecture, offering a powerful, general framework for decoding the temporal dimension of complex systems.

\section*{Methodological Framework}

\begin{figure}
	\centering
	\includegraphics[scale=0.9]{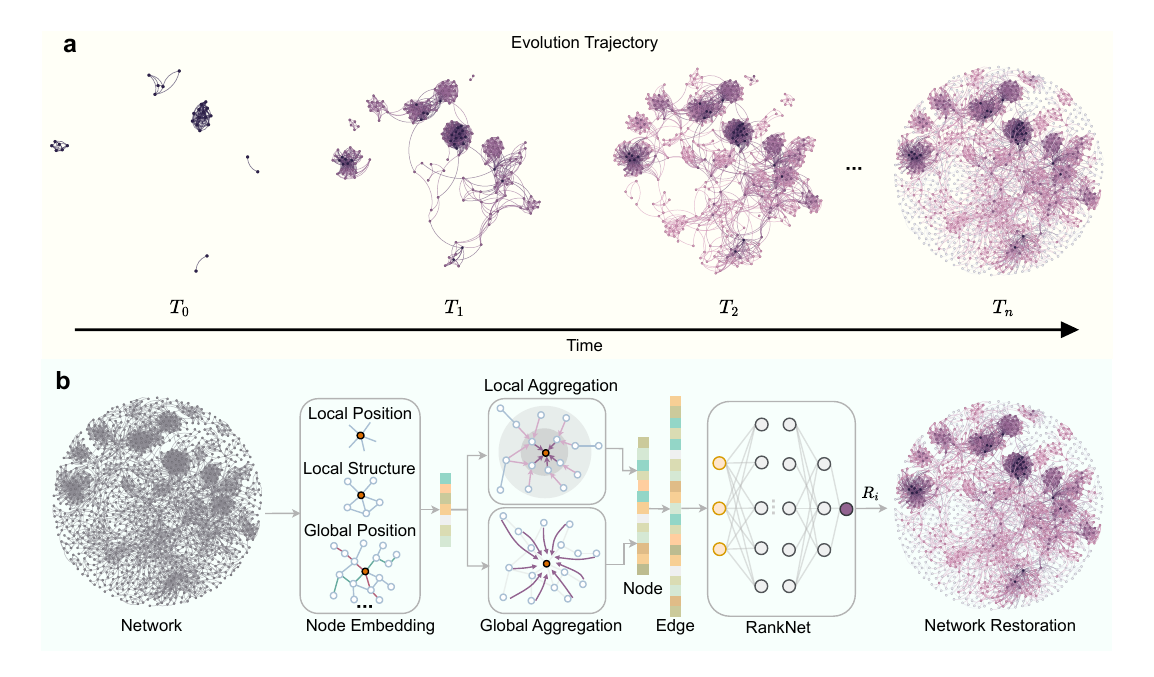}
	\caption{
		\textbf{Network evolution and machine learning framework for restoring structure evolution.} 
		(a) Diagram of network formation, where new edges are added at each time step, with the goal of restoring the edge formation sequence from the final network structure alone. (b) Overview of the machine learning framework for inferring network evolution. Node embeddings are initialized using topological features like degree, clustering coefficient, and betweenness (see Supplementary Sec.~1.1), then processed by GNN and transformer modules to capture local and global structural patterns. Edge embeddings, formed by combining endpoint node embeddings, are fed into a RankNet model to predict relative edge formation times ($R_i$), enabling restoration of the network’s evolutionary trajectory.}
	\label{demo}
\end{figure}

This study uses machine learning to restore the order in which connections form in evolving networks, as shown in Fig.~\ref{demo}. We adopt a two-step approach. First, for networks where we know part of the connection history, we create a supervised learning model that combines the network’s structure with its known past to predict how it formed. Second, for networks where only the final structure is available, we transfer the model trained in the first step to similar networks to figure out the target network’s formation process.

In the first step, we map the edges of a network with partial evolution data into a low-dimensional space. We then create pairs of these edges to train a machine learning model to predict their relative formation order, using only pairs with known formation sequences derived from the network’s history. Our approach uses a hybrid model, GNN-Transformer-Ranknet (GTR), combining GNNs and transformers to infer the full edge formation sequence (see Fig.~\ref{demo}b, Materials and Methods, and Supplementary Sec.~1 for details). 

In the second step, for networks without any evolution data, we exploit the model’s transferable capability to restore their formation history.
The target network is embedded into a low-dimensional space using the same method as in the first step. The pretrained GTR model then predicts the edge formation sequence. 
A major challenge in cross-network prediction is aligning the embedding spaces of source and target networks, under the assumption that their growth mechanisms are similar.
Our results rely on an embedding approach that extracts hierarchical, isomorphism-invariant topological features (see Supplementary Sec.~1.1.1), which are consistent across networks and enable effective transfer. This allows our model to accurately restore the evolution of target networks without any prior temporal data.

\section*{Restoration of Network Dynamics from Partial Connection History}
\subsection*{High Accuracy Restoration}
\begin{figure}
\centering
\includegraphics[scale=0.7]{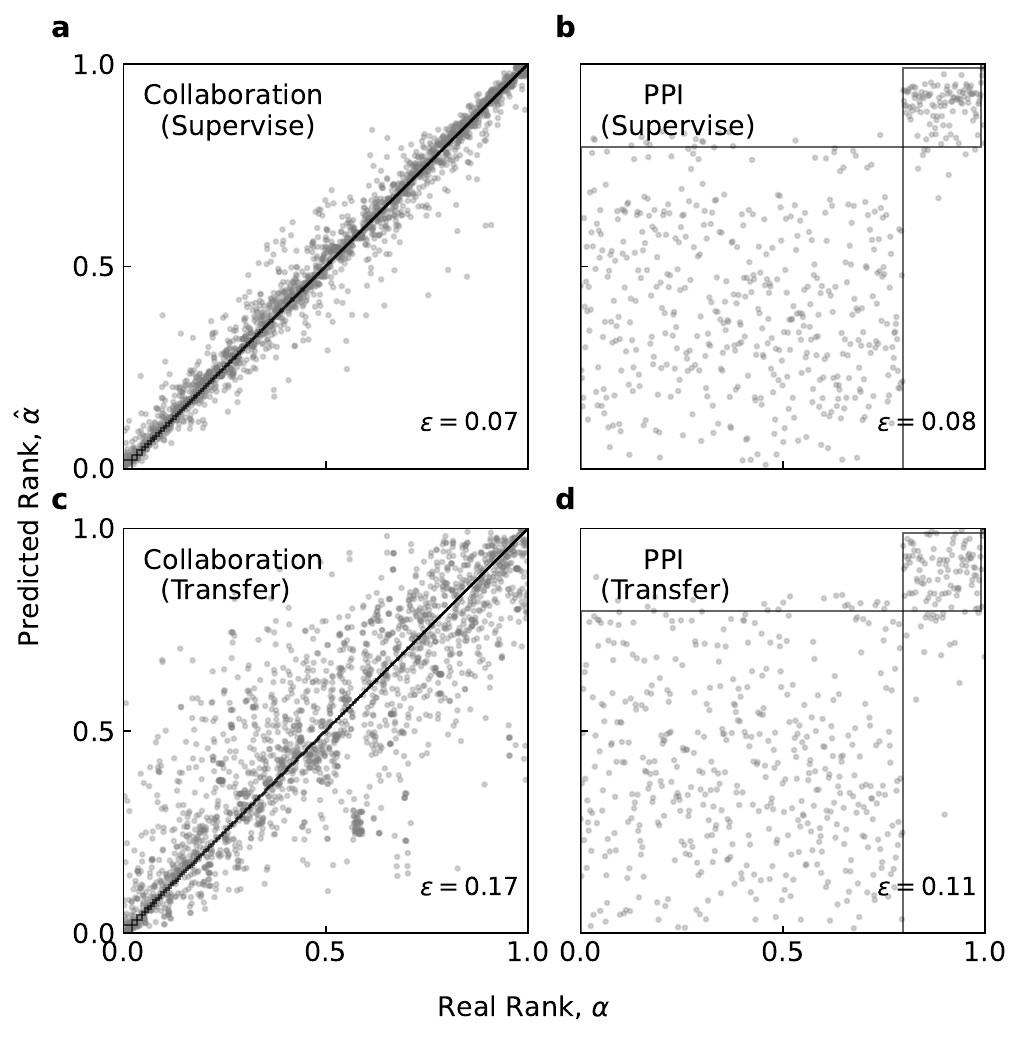}
\caption{%
\textbf{Comparison between true and predicted edge formation times.}
Results of collaboration and PPI networks are shown in the left and right columns, respectively. 
Panels (a,b) correspond to the \textit{supervised setting}, where   models are trained and tested on the same network: 
(a) Collaboration (Fluctuations$\rightarrow$Fluctuations); 
(b) PPI (Fruit fly$\rightarrow$Fruit fly). 
Panels (c,d) correspond to the \textit{cross-network transfer} setting, where models are trained and tested on different networks: 
(c) Collaboration (Chaos$\rightarrow$Fluctuations); 
(d) PPI (Human$\rightarrow$Fruit fly). 
Rectangles highlight edges that are formed within the same snapshot in the true formation times. 
Here, $\alpha$ and $\hat{\alpha}$ denote the ground-truth and predicted formation time, respectively, both normalized by the total number of edges $E$. 
The prediction error is quantified by $\varepsilon$ (see definition in Materials and Methods). Results for additional networks are reported in Table~\ref{tltab_error}.
The GTR model is used for all results.}
\label{scatterplot}
\end{figure}

To test our machine learning framework, we ran experiments on 16 real-world networks with time-stamped snapshots, including protein–protein interaction (PPI), scientific collaboration, traffic, animal interaction, and world trade networks (see Supplementary Sec.~2 for details). We defined each edge’s formation time as the first snapshot it appears in and created edge pairs with known formation orders for training and testing. Prediction accuracy ($x$) is the fraction of correctly ordered edge pairs in the test set. We compared our GTR model to the CPNN model from Ref.~\cite{wan24} and other baseline models, with results shown in Table~\ref{sl}. GTR outperformed others in 12 of the 16 networks, with accuracy gains over 10\% in 9 cases, notably reaching 90\% on the PPI (Fruit fly) network—a 20\% improvement over CPNN. Fig.~\ref{compare}a–f shows GTR’s consistent superiority across various training ratios and network types (see Supplementary Fig.~S3 for more). Accuracy climbs above 80\% as training data increases, leveling off near a 5\% training ratio. Even simpler versions of our model, R (RankNet alone) and GR (GNN with RankNet), beat CPNN by over 5\% in at least 9 networks, underscoring the strength of our framework and its isomorphism-invariant features.
Figure \ref{scatterplot}a and \ref{scatterplot}b illustrate a clear comparison between the actual and predicted formation times of edges.

In this work, we mainly use pairwise accuracy $x$ to assess restoration quality, as coarse-grained temporal data in real-world networks complicates direct comparison between ground-truth and restored sequences. Theoretical results from Ref. (16) show that $x$ and restoration error $\varepsilon^{theory}$ are closely related, with $\varepsilon^{theory}$ approaching zero when $x$ exceeds 0.5, following the relation 
$
\varepsilon^{theory}=\frac{\sqrt{x(1 - x)}}{2x - 1}\frac{1}{\sqrt{E}}
$~\cite{wan24}
where $E$ is the total number of edges. Our experiments achieve $x$ well above this threshold, confirming the reliability of the restored network evolution trajectory for downstream applications.

\subsection*{Few-Shot Learning}
Noticing that GTR’s accuracy stabilizes with minimal training data, we tested its performance using few supervised samples. Fig.~\ref{compare}g illustrates how accuracy shifts with the number of training edge samples. Across four representative networks, accuracy beats random guessing ($x > 50\%$) with just over 10 edges. By 64 edges, three networks reach over 80\% accuracy. More comparisons across additional networks and models are available in Supplementary Fig.~S4. These results show GTR can accurately restore network evolution with only a handful of time-labeled edges, highlighting its strong potential for few-shot learning.

\subsection*{Low Spatiotemporal Complexity}
Beyond its strong predictive accuracy, our GTR model excels in computational efficiency and simplicity. As shown in Fig.~\ref{compare}h, GTR trains about 120 times faster than CPNN while matching its accuracy in the PPI (Fungi) network (see Supplementary Fig.~S5 for more results). Additionally, GTR uses only 20,081 parameters—roughly a third of CPNN’s 56,017 (Fig.~\ref{compare}i)—making it highly efficient and scalable.

\section*{Transfer-Based Restoration of  Network Dynamics from Static Structure}
Using a supervised-learning model trained on a source network with partial historical data, we exploit its cross-network transferability to restore the evolution of a target network without any temporal information. Fig.~\ref{tl}a shows the transfer accuracy matrix using isomorphism-invariant topological features as initial node embeddings. Entries near the diagonal achieve around 70\% accuracy, well above the 50\% random baseline, but accuracy drops further from the diagonal, indicating better transfer within network categories (e.g., PPI or scientific collaboration) than across them. The model trained on the World Trade Web performs poorly, at near-random levels. Similarly, using random-walk, non-random-walk, or matrix-factorization embeddings results in transfer accuracy around 50\%, as shown in Fig.~\ref{tl}f. Comparing models with isomorphism-invariant features (Fig.~\ref{tl}b), all achieve over 60\% accuracy, with GTR surpassing 80\%. Table~\ref{tltab} summarizes average accuracy when training on one network and testing within the same network category, with GTR leading in 12 of 15 networks, showcasing its strong transferability across networks.
Figures \ref{scatterplot}c and \ref{scatterplot}d present a clear comparison between the actual and predicted edge formation times. The transfer results are comparable to those in the supervised setting.
Table 2 provides quantitative support for this observation, demonstrating that prediction errors in transfer settings are notably lower than those observed in random baselines and closely align with supervised-learning performance levels, thereby confirming the model's effective generalization of temporal restoration across diverse network types.

To explore why transfer performance varies, we used t-SNE~\cite{hge02} to project isomorphism-invariant features as node embeddings of five network categories onto a 2D plane (Fig.~\ref{tl}d). Nodes from the same category, like “Phase Transitions” and “Fluctuations” collaboration networks, cluster closely, while World Trade Web nodes form a distinct, isolated group. In contrast, random-walk embeddings (Fig.~\ref{tl}e) create tight clusters for each network, but these are far apart, even within categories. This highlights the importance of aligning node embedding distributions for effective cross-network transferability. We introduced a Spearman correlation–based metric to measure structural distance between networks (see Supplementary Sec.~3). Fig.~\ref{tl}c shows transfer accuracy plotted against this distance, revealing a strong negative correlation (slope: $-1.86 \pm 0.24$), confirming that structurally similar networks yield better transfer performance and offering a way to select optimal source networks for training.

\section*{Application to Restore \textit{Drosophila} Brain Network}
To test the cross-network transferability of our machine learning model, we applied it to restore the detailed evolutionary trajectory of the \textit{Drosophila} brain network, which includes about 2.6 million neural connections across 140,000 neurons \cite{dor24,sch24}. We trained the model on a network most similar to \textit{Drosophila}, based on the distance metric in Supplementary Sec.~3, then used it to estimate the formation time of each connection in the \textit{Drosophila} brain. We focused further analysis on the olfactory circuit subnetwork, crucial for behaviors like foraging, courtship, mating, and aggression~\cite{bar06,suh04,ram15}. The predicted evolutionary trajectory of the \textit{Drosophila} brain network can be seen in Supplementary Sec.~9 and Supplementary Video.

Fig.~\ref{fly} shows the predicted formation time distribution of the top ten most common neural connections in the olfactory circuit, using the PPI (worm) network as the training source (details in Supplementary Sec.~8). The results highlight clear differences in formation times: connections linked to sex pheromone sensitivity and gender recognition \cite{dat08} form early, as do those aiding food detection \cite{sil11}. In contrast, connections tied to complex functions, like dialect learning in social settings \cite{kac19} or detecting specific odors during foraging \cite{lud05}, appear later. More details on these connections and their functions are in Supplementary Sec.~2.6. A schematic of the glomeruli’s spatial arrangement in the antennal lobe is included in Fig.~\ref{fly}’s inset.

These findings reveal a correlation between the predicted forming time and functional essentiality of the neural connections. Early-forming connections support important survival and reproductive behaviors, while later-forming ones are tied to adaptive behaviors in complex environments. This hierarchy provides new insights into the \textit{Drosophila} olfactory system’s behavioral optimization and evolution. Remarkably, our model relied only on network structure and isomorphism-invariant features, without any prior \textit{Drosophila}-specific biological knowledge.

\section*{Discussion and Conclusion}

Understanding how complex networks evolve is central to many disciplines, from neuroscience and systems biology to social and information sciences. Here, we developed a transferable machine-learning framework that combines GNNs and transformers to restore network evolution from static structure alone. By jointly capturing local and global topological features, our model achieves state-of-the-art accuracy, computational efficiency, and scalability across diverse empirical networks. Leveraging isomorphism-invariant topological descriptors, it aligns structural embeddings across systems, enabling robust cross-network transferability even in the absence of temporal data.

The ability to infer historical dynamics from static configurations opens a new analytical dimension in network science. Our approach demonstrates that much of the temporal information governing the evolution of complex systems remains implicitly encoded within their topology. This insight has direct implications for restoring developmental and functional pathways in biological and technological networks alike. In particular, applying the framework to the \emph{Drosophila} brain revealed a strong correlation between predicted connection age and behavioral essentiality—early-forming connections supporting core functions such as mating and foraging, while later-forming ones underpin more specialized, adaptive behaviors. These results suggest that evolutionary and functional hierarchies may be recoverable from structure alone.

Despite these advances, several challenges remain. First, our framework assumes that once a connection forms, it persists, whereas many real-world networks—such as social, ecological, or communication systems—undergo continuous rewiring. Extending the model to account for edge deletion and node turnover would allow for truly dynamic restorations. Second, while the framework operates without domain-specific information, incorporating contextual knowledge (e.g., biological constraints or semantic attributes) could enhance interpretability and accuracy. Finally, applying this approach to multiscale or multilayer systems represents an exciting direction for future research, potentially bridging micro-level processes and macroscopic structural evolution.

In summary, our work establishes a general, efficient, and transferable approach for restoring the temporal evolution of complex networks from static data. By mapping the full formation history of over 2.6 million neural connections in the \emph{Drosophila} brain, we uncover fundamental links between structural development and function. Beyond neuroscience, this methodology provides a foundation for decoding the hidden temporal dimension of complex systems across biology, ecology, social dynamics, and information networks, offering a new lens through which to understand how structure, function, and evolution intertwine. Finally, this cross-domain universality suggests that the same underlying structural invariants encode evolutionary memory across systems.

\section*{Materials and Methods}
\subsection*{Embedding Methods} \label{emb}
To model the evolution of complex networks with machine learning, we first need suitable inputs. We evaluated seven node embedding methods: 1. Isomorphism-invariant topological features, capturing 20 standard metrics like degree and clustering coefficient. 2. Random-walk-based approaches: Node2Vec~\cite{gro16} and DeepWalk~\cite{per14}. 3. Non-random-walk methods: LINE~\cite{tan15} and SDNE~\cite{wan16}. 4. Matrix-factorization techniques: GraRep~\cite{cao15} and HOPE~\cite{oum16}. See Supplementary Sec.~1.1 for more details.

\subsection*{Machine Learning Model} \label{GTR}
A key task in our study is predicting the relative formation order of edge pairs. We developed the GTR architecture, which combines local and global structural analysis. The GNN-Transformer (GT) module generates high-quality node embeddings by encoding both local and global network features \cite{ram22}, while the RankNet (R) module ranks edge pairs by formation order \cite{bur05}. The GT module takes initial node embeddings and network topology as input, producing updated node embeddings. For each edge, we combine the embeddings of its endpoints to form an edge embedding, which the R module uses to predict relative formation order. During inference, the R module assigns a score to each edge, allowing us to sort them into a complete formation sequence. Unlike the method in Ref.~\cite{wan24}, which has a test-time complexity of $O(N^2)$ due to exhaustive pairwise comparisons (where $N$ is the number of edges), our approach reduces this to $O(N)$, greatly enhancing efficiency and scalability for large networks. Additional model details are in Supplementary Sec.~1.

\subsection*{Prediction Error}\label{error}
Let $\alpha_i$ denote the position of edge $i$ in the ground-truth edge sequence (e.g., $\alpha_i = i$, with larger $\alpha_i$ indicating that edge $i$ was generated later), and let $\hat{\alpha}_i$ denote its corresponding position in the predicted sequence produced by our machine learning model. Define $\alpha = (\alpha_1, \alpha_2, \dots, \alpha_E)$ as the ground-truth formation times and $\hat{\alpha} = (\hat{\alpha}_1, \hat{\alpha}_2, \dots, \hat{\alpha}_E)$ as the predicted formation times, both normalized by the total number of edges $E$. 
The prediction error is quantified using the root-mean-squared error:
$\varepsilon = \sqrt{\frac{1}{E} \sum_{i=1}^{E} d_i^2 }$,
where $d_i = \alpha_i - \hat{\alpha}_i$ represents the deviation of edge $i$. 
In the ground-truth sequence $\alpha$, edges might be generated at the same time, leading to a coarse-grained representation where the exact order of edges within a snapshot cannot be distinguished. To account for this, we define intervals for each snapshot that span all edges generated at the same time, and use the midpoint of this interval as the representative formation time, denoted by $\alpha_i^\mathrm{m}$, for all edges in that snapshot. 
If the predicted formation time $\hat{\alpha}_i$ falls within the interval defined for an edge $\alpha_i$, we consider the prediction to be correct, and thus set $d_i = 0$. If $\hat{\alpha}_i$ falls outside the interval, the error is calculated as the difference between the predicted formation time and the midpoint of the interval, i.e., $d_i = \alpha_i^\mathrm{m} - \hat{\alpha}_i$.



\begin{figure}
\centering
\includegraphics[scale=0.25]{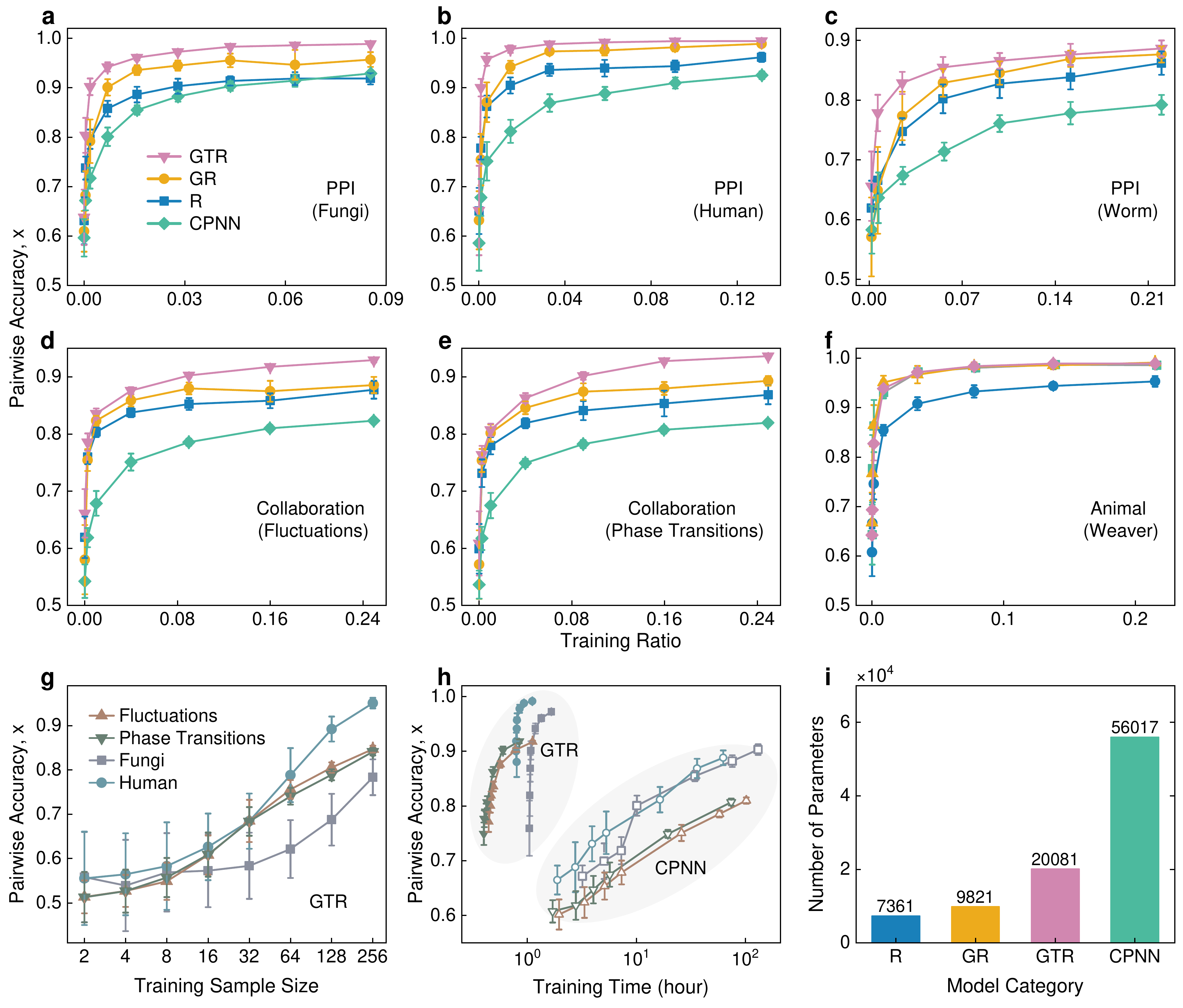}
\end{figure}

\begin{figure}
\centering
\caption{\textbf{Comparison of different models in terms of accuracy, computational cost, and model size.} 
(a–f) Pairwise edge prediction accuracy on the test set under different training ratios of edge pairs. Error bars represent standard deviations over 50 runs. 
Green: CPNN; Blue: R; Orange: GR; Pink: GTR. 
Results are shown for three representative network categories: protein–protein interaction network, scientific collaboration network, and animal interaction network. See Supplementary Fig.~S3 for results on additional network categories. 
(g) Few-shot learning performance of the GTR model. Accuracy across the number of training edges on four representative networks. See Supplementary Fig.~S4 for results on additional networks and models. 
(h) Comparison of accuracy and training time between GTR and CPNN on four representative networks, as shown in (g).  Experiments were conducted on a workstation equipped with an Intel i7-13700KF CPU, 80GB RAM, and an NVIDIA RTX 4070Ti GPU (see Supplementary Fig.~S5 for results on additional networks and models).
(i) Model parameter size comparison of the four models.
}\label{compare}
\end{figure}

\begin{figure}
\centering
\includegraphics[scale=0.65]{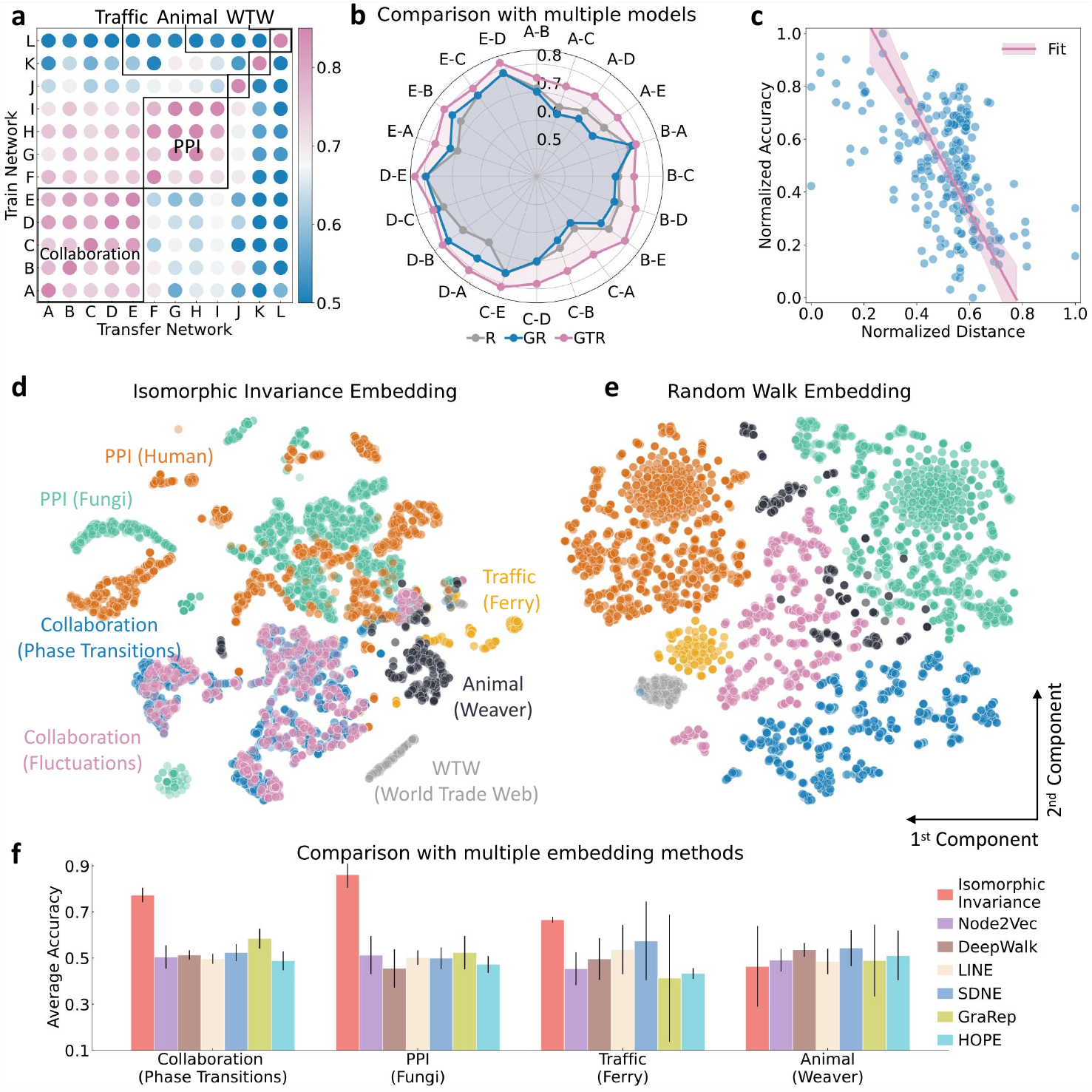}
\end{figure}

\begin{figure}
\centering
\caption{\textbf{Results of cross-network transferability.}
(a) Accuracy matrix of cross-network transferability for the GTR model, using isomorphism-invariant features as initial node embeddings. The color intensity of each matrix cell reflects the transfer accuracy between source and target networks.
The Y-axis denotes the source network (where the model is trained), and the X-axis denotes the target network (where the model is tested). 
Networks are clustered by category, and blocks from the same category are marked with black squares and corresponding labels. 
Network labels: A: Thermodynamics, B: Complex networks, C: Phase transitions, D: Fluctuations, E: Chaos, F: Worm, G: Fruit fly, H: Human, I: Fungi, J: Ferry, K: Weaver, L: World Trade Web. Results of the synthetic networks, please refer to Supplementary Sec.~7.
(b) Comparison of transfer accuracy across three models with node embeddings initialized by isomorphism-invariant features. Each $X\text{-}Y$ pair indicates a model trained on network $X$ and transferred to network $Y$.  
(c) Scatter plot of transfer accuracy versus cross-network distance (see Supplementary Sec.~3 for definition) using the GTR model. The pink line shows a linear fit with a slope of $-1.86\pm0.24$, and the light pink shaded area indicates the 95\% confidence interval. 
Both accuracy and distance values are Min-Max normalized: $x_{norm} = (x - x_{min}) / (x_{max} - x_{min})$.
(d–e) Two-dimensional projection of node embeddings via t-SNE~\cite{hge02}: (d) isomorphism-invariant embedding; (e) random walk embedding. Each dot represents a node, colored by its network type. Labels indicate the corresponding network names. 
(f) Comparison of average transfer accuracy across seven initialization methods of node embeddings using the GTR model. 
Average accuracy refers to the mean pairwise edge prediction accuracy of a model trained on one source network and transferred to other target networks within the same category. 
Details of seven embedding methods please refer to Materials and Methods and Supplementary Sec.~1.1.
}
\label{tl}
\end{figure}

\begin{figure}
\centering
\includegraphics[scale=1.2]{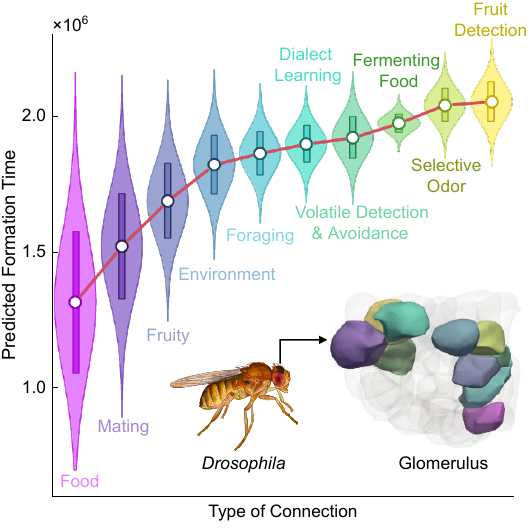}
\caption{\textbf{Predicted connection formation time and functional organization of the \textit{Drosophila} olfactory circuit}. Violin plots of predicted formation times of neural connections, grouped by functional connection type in the \textit{Drosophila} olfactory circuit. Each color denotes a distinct category of connections (text labels), white dots mark the mean formation time, and vertical bars indicate the standard error. Inset: spatial map of glomeruli where neural connections form, with highlighted regions corresponding to the same connection categories and gray areas indicating other glomeruli. Details please refer to Supplementary Sec.~2.6. The \textit{Drosophila} illustration is from the DataBase Center for Life Science.}

\label{fly}
\end{figure}


\newpage
\begin{table}
\centering
\caption{\textbf{Performance of different models in the supervised learning.} 
Pairwise edge prediction accuracy of different models for various networks, using isomorphism-invariant features as the initial node embedding. 
CPNN refers to the model proposed in~\cite{wan24}. 
R denotes RankNet alone, GR combines GNN and RankNet, while GTR integrates GNN, Transformer, and RankNet (details in Supplementary Sec.~1). 
Each value represents the mean accuracy over 50 runs, with standard deviation shown in parentheses. 
The best-performing results are highlighted in bold. Twenty percent of each network’s edges are used for training.}
\resizebox{\textwidth}{!}{
\begin{tabular}{clcccc}
\hline
\textbf{Network Category} & \textbf{Network Name} & \textbf{CPNN} & \textbf{R} & \textbf{GR} & \textbf{GTR} \\
\hline
& Bacteria   & 0.87 (0.02) & 0.88 (0.02) & 0.87 (0.02) & \textbf{0.95 (0.02)} \\ 
& Fruit fly  & 0.70 (0.04) & 0.77 (0.04) & 0.77 (0.07) & \textbf{0.90 (0.03)} \\ 	
\textbf{Protein-Protein Interaction}& Fungi      & 0.80 (0.02) & 0.84 (0.02) & 0.90 (0.02) & \textbf{0.94 (0.01)} \\
& Human      & 0.81 (0.02) & 0.90 (0.02) & 0.94 (0.01) & \textbf{0.98 (0.01)} \\ 
& Worm       & 0.67 (0.01) & 0.75 (0.02) & 0.77 (0.04) & \textbf{0.83 (0.02)} \\ 
\hline
& Chaos             & 0.74 (0.01) & 0.85 (0.01) & 0.87 (0.01) & \textbf{0.89 (0.01)} \\
& Complex networks  & 0.68 (0.02) & 0.79 (0.02) & 0.78 (0.03) & \textbf{0.82 (0.02)} \\ 
\textbf{Collaboration}& Fluctuations      & 0.75 (0.01) & 0.84 (0.01) & 0.86 (0.01) & \textbf{0.88 (0.01)} \\ 
& Phase transitions & 0.75 (0.01) & 0.82 (0.01) & 0.85 (0.01) & \textbf{0.86 (0.01)} \\ 
& Thermodynamics    & 0.68 (0.04) & 0.73 (0.03) & 0.78 (0.02) & \textbf{0.80 (0.03)} \\ 
\hline
& Aircraft & \textbf{0.88 (0.04)} & 0.87 (0.04) & 0.84 (0.05) & 0.87 (0.04) \\ 
\textbf{Traffic}& Coach    & 0.80 (0.09) & 0.69 (0.09) & 0.80 (0.05) & \textbf{0.85 (0.09)} \\
& Ferry    & \textbf{0.78 (0.04)} & 0.76 (0.05) & 0.76 (0.04) & 0.75 (0.05) \\ 
\hline
\multirow{2}{*}{\textbf{Animal}} 
& Ants   & 0.78 (0.03) & 0.85 (0.03) & 0.75 (0.04) & \textbf{0.86 (0.02)} \\ 
& Weaver & \textbf{0.97 (0.01)} & 0.91 (0.01) & \textbf{0.97 (0.02)} & \textbf{0.97 (0.01)} \\
\hline
\textbf{World Trade Web}
& WTW    & \textbf{0.89 (0.01)} & 0.85 (0.01) & 0.86 (0.01) & 0.88 (0.01) \\
\hline
\end{tabular}
}
\label{sl}
\end{table}

\begin{table*}[t!]
\centering
\caption{\textbf{Prediction error $\varepsilon$ under different settings.}
\textit{Supervise} refers to within-network supervised learning,
\textit{Transfer} denotes cross-network transfer,
and \textit{Random} indicates the error obtained by randomly shuffling the transfer-predicted sequence before comparison with the ground truth.
Results are produced using the GTR model. Each value represents the mean prediction error over 50 independent runs, 
with standard deviations shown in parentheses.}
\begin{tabular}{clccc}
\hline
\textbf{Network Category} & \textbf{Network Name} & \textbf{Supervise} & \textbf{Transfer} & \textbf{Random} \\
\hline
& Bacteria & 0.06 (0.01) & 0.10 (0.01) & 0.19 (0.01) \\
& Fruit fly & 0.08 (0.01) & 0.11 (0.01) & 0.30 (0.01) \\
\textbf{Protein-Protein Interaction}& Fungi & 0.06 (0.01) & 0.14 (0.01) & 0.22 (0.01) \\
& Human & 0.05 (0.01) & 0.15 (0.02) & 0.31 (0.01) \\
& Worm & 0.17 (0.01) & 0.21 (0.01) & 0.38 (0.01) \\
\hline
& Chaos & 0.08 (0.01) & 0.16 (0.01) & 0.41 (0.01) \\
& Complex networks & 0.10 (0.01) & 0.18 (0.01) & 0.41 (0.01) \\
\textbf{Collaboration} & Fluctuations & 0.07 (0.01) & 0.17 (0.01) & 0.41 (0.01) \\
& Phase transitions & 0.07 (0.01) & 0.20 (0.01) & 0.41 (0.01) \\
& Thermodynamics & 0.11 (0.01) & 0.19 (0.02) & 0.41 (0.01) \\
\hline
& Aircraft & 0.09 (0.02) & 0.12 (0.02) & 0.21 (0.02) \\
\textbf{Traffic}& Coach & 0.03 (0.01) & 0.05 (0.01) & 0.08 (0.01) \\
& Ferry & 0.12 (0.02) & 0.17 (0.01) & 0.23 (0.01) \\
\hline
\multirow{2}{*}{\textbf{Animal}} 
& Ants & 0.07 (0.01) & 0.09 (0.01) & 0.13 (0.01) \\
& Weaver & 0.04 (0.01) & 0.32 (0.03) & 0.40 (0.01) \\
\hline
\textbf{World Trade Web}
& WTW & 0.11 (0.01) & 0.27 (0.05) & 0.29 (0.01) \\
\hline
\end{tabular}
\label{tltab_error}
\end{table*}

\begin{table*}[t!]
\centering
\caption{\textbf{Performance of different models in cross-network transferability.}
Pairwise edge prediction accuracy of different models for various networks, using isomorphism-invariant features as the initial node embedding. 
Models are trained on a single source network and then transferred to other target networks within the same network category. 
Values represent the mean accuracy over 50 runs, and standard deviations are shown in parentheses. 
The best-performing results are highlighted in bold. Fifty percent of each network’s edges are used for training. 
}
\begin{tabular}{clccc}
\hline
\textbf{Network Category} & \textbf{Network Name} & \textbf{R} & \textbf{GR} & \textbf{GTR}  \\
\hline
& Bacteria & 0.64 (0.10) & 0.64 (0.07) & \textbf{0.80 (0.07)}  \\
& Fruit fly & 0.71 (0.07) & 0.69 (0.07) & \textbf{0.80 (0.09)} \\
\textbf{Protein-Protein Interaction} & Fungi & 0.66 (0.10) & 0.70 (0.10) & \textbf{0.86 (0.06)} \\
& Human & 0.63 (0.08) & 0.65 (0.06) & \textbf{0.87 (0.07)} \\
& Worm & 0.69 (0.07) & 0.67 (0.06) & \textbf{0.74 (0.03)} \\
\hline
& Chaos & 0.74 (0.04) & 0.76 (0.03) & \textbf{0.80 (0.02)} \\
& Complex networks & 0.72 (0.02) & 0.70 (0.04) & \textbf{0.77 (0.02)} \\
\textbf{Collaboration} & Fluctuations & 0.74 (0.05) & 0.78 (0.02) & \textbf{0.81 (0.02)} \\
& Phase transitions & 0.69 (0.06) & 0.68 (0.07) & \textbf{0.77 (0.03)} \\
& Thermodynamics & 0.69 (0.02) & 0.66 (0.03) & \textbf{0.75 (0.01)} \\
\hline
& Aircraft & \textbf{0.61 (0.03)} & \textbf{0.61 (0.01)} & 0.58 (0.01) \\
\textbf{Traffic} & Coach & 0.61 (0.05) & 0.54 (0.01) & \textbf{0.67 (0.09)} \\
& Ferry & 0.65 (0.21) & 0.55 (0.23) & \textbf{0.67 (0.01)} \\
\hline
\multirow{2}{*}{\textbf{Animal}} 
& Ants & 0.49 (0.01) & \textbf{0.53 (0.04)} & 0.49 (0.03) \\
& Weaver & \textbf{0.55 (0.19)} & 0.47 (0.17) & 0.46 (0.18) \\
\hline
\end{tabular}
\label{tltab}
\end{table*}


\clearpage 

%
\bibliography{references} 
\bibliographystyle{sciencemag}

\end{document}